\documentclass[twocolumn,secnumarabic,,amsmath,amssymb, nobibnotes, aps, prd, superscriptaddress]{revtex4-1}

\usepackage{amsmath}
\usepackage{amssymb}
\usepackage{graphicx}
\usepackage{dcolumn}
\usepackage{floatrow}
\usepackage{xcolor}

\begin{document}
\title{Plasmons and screening in finite-bandwidth 2D electron gas}
\author{Kaveh Khaliji}
\affiliation{Department of Electrical and Computer Engineering, University of Minnesota, Minneapolis, Minnesota 55455, USA}
\author{Tobias Stauber}
\affiliation{Materials Science Factory, Instituto de Ciencias de Materiales de Madrid, CSIC, E-28049, Madrid, Spain}
\author{Tony Low}
\email{tlow@umn.edu}
\affiliation{Department of Electrical and Computer Engineering, University of Minnesota, Minneapolis, Minnesota 55455, USA}

\begin{abstract}
The dynamical and nonlocal dielectric function of a two-dimensional electron gas (2DEG) with finite energy bandwidth is computed within random-phase approximation. For large bandwidth, the plasmon dispersion has two separate branches at small and large momenta. The large momenta branch exhibits negative quasi-flat dispersion. The two branches merge with decreasing bandwidth. We discuss how the maximum energy plasmon mode which resides at energies larger than all particle-hole continuum can potentially open a route to low-loss plasmons. Moreover, we discuss the bandwidth effects on the static screening of the charged and magnetic impurities. 
\end{abstract}

\maketitle

\noindent \textit{Introduction} -- Since early 1950s, there has been on-going interests in the theory of two-dimensional (2D) electronic systems. This has been partly fueled by the development of metal-oxide-semiconductor transistors, where the surface accumulated charge layers formed in the device behaves as a two-dimensional electron gas (2DEG) \citep{bardeen1948, ando1982, streetman1995}. Isolation of atomically thin layers from bulk parent materials \citep{novoselov2004}, has also boosted this interest, where the theory of 2DEG has served as point of reference for the elementary electronic properties of graphene and other 2D systems \citep{sarma2011, wunsch2006, hwang2007, scholz2012, low2014}.\\

The dynamical 2D polarizability, which describes the screening of the Coulomb potential is vital for understanding many physical properties in 2D systems. For example, dynamical screening governs the elementary excitation spectra and gives the collective modes dispersion. Moreover, the static screening dictates the magnetic response via control over exchange interactions between localized magnetic impurities embedded in the 2D metal, or determines the transport properties through screened Coulomb scattering by charged impurities \cite{giuliani2005, mahan2013}.\\

With the 2DEG polarizability now a textbook example covered in many-body physics \citep{giuliani2005}, we ask a seemingly simple question: how does the polarizability function gets modified if we restrict the parabolic energy dispersion with an energy cut off? This was motivated by recent developments in several materials system exhibiting isolated electronic bands with finite bandwidths. This includes TaS$_{2}$ or NbS$_{2}$ single layers in 2H phase \citep{gjerding2017}, or Si(111):X in $\alpha-\sqrt{3}\times\sqrt{3}$ phase, constructed by group IV adatom X adsorbed on silicon surface (111) \citep{schuwalow2010, hansmann2013, wang2015}. We show a finite bandwidth allows for plasmon modes of quasi-flat dispersion and large momenta to emerge. The finite bandwidth 2DEG (FBW-2DEG) can also potentially support low-loss plasmon modes immune to elastic or inelastic scattering-assisted Landau damping (dissipation via electron-hole pair excitation). We show the static polarizability approaches a constant value at large momenta limit and the value can be tuned with bandwidth and Fermi energy. We find in FBW-2DEG, static Friedel oscillations due to a charged impurity can be strongly damped as the bandwidth shrinks. However, the induced spin density due a magnetic impurity, has a large distance oscillatory decay similar to the 2DEG, and is not affected by the bandwidth.\\

\noindent \textit{Dynamic Polarizability and Plasmon Modes} -- We assume an electronic energy dispersion of the form: $E_{\bf{k}} = \hbar^2 k^2 /2m$ for $k \leq k_c$ and $E_{\bf{k}} = E_c$ otherwise. The Fermi energy and momentum are denoted with $E_{F}$ and $k_{F}$, respectively. Accordingly, we introduce: $\bar{E}_{c} = E_{c}/E_{F}$ and $\bar{k}_{c} = k_{c}/k_{F}$. In Fig. 1(a) the main symbols pertaining to FBW-2DEG electron energy dispersion are illustrated. In this work we assume $\bar{k}_{c}>1$.\\

The dielectric function in the random phase approximation reads as:
\begin{equation}
\epsilon({\bf{q}}\,, \omega) = 1 + \frac{e^2}{2\epsilon_{0}\kappa q} P({\bf{q}}\,,\omega)
\label{eps_RPA}
\end{equation}
with the Lindhard polarizability given as:
\begin{equation}
P({\bf{q}}\,, \omega) = -\frac{2}{\mathcal{S}} \sum _{\bf{k}}\frac{f_{\bf{k}}-f_{{\bf{k}}+{\bf{q}}}}{E_{{\bf{k}}} - E_{{\bf{k}}+{\bf{q}}} + \hbar\omega +i\eta},
\label{pol_lindhard}
\end{equation}
where, $\mathcal{S}$ is the sample area and $\kappa$ is the background dielectric constant. We use $m/m_{0} = 0.5$, $E_{F} = 0.1$\,eV, $\kappa = 2.5$, unless denoted otherwise. We assume zero temperature. The undamped plasmon modes correspond to zeros of the dielectric function. For finite broadening, where $\operatorname{Im}\,P \neq 0$, the loss function, defined as: $L({\bf{q}}\,, \omega) = - \operatorname{Im} \left( 1/\epsilon({\bf{q}}\,, \omega)\right)$ is used to visualize the low-loss plasmon modes which appear as sharp peaks \citep{stauber2014}.\\

\begin{figure*}
\centering
\includegraphics[width=\linewidth]{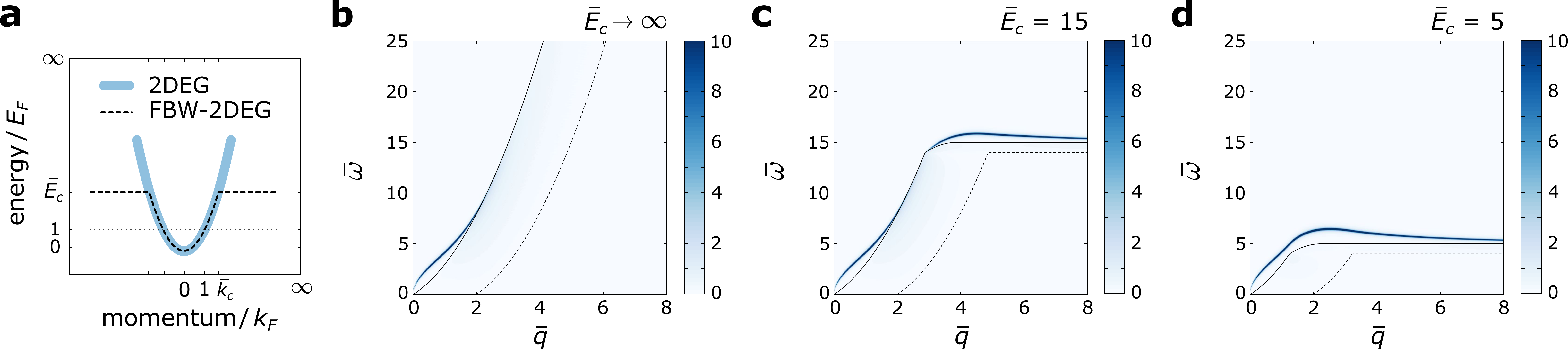}
\caption{(a) The 2DEG and FBW-2DEG electron energy dispersions. The dotted line is the Fermi energy. The loss function for (b) 2DEG, (c) $\bar{E}_{c} = 15$, and (d) $\bar{E}_{c} = 5$. The solid and dashed black lines denote upper and lower boundaries of the PHC, respectively. We take $\eta = 5$\,meV.}
\label{fig1}
\end{figure*}

For 2DEG, i.e. when $E_c \rightarrow \infty$, the plasmon dispersion has the analytic form $(\eta \rightarrow 0)$ \citep{czachor1982}:
\begin{equation}
\begin{split}
\bar{\omega}^{2} = & 2\bar{q}_{\textsf{TF}}\bar{q}\left(1+\frac{\bar{q}}{\bar{q}_{\textsf{TF}}}\right)^2 \\
& \times \left(1+\frac{\bar{q}^3}{2\,\bar{q}_{\textsf{TF}}}+\frac{\bar{q}^4}{4\,\bar{q}_{\textsf{TF}}^2}\right)\bigg / \left(1+\frac{\bar{q}}{2\bar{q}_{\textsf{TF}}}\right),
\end{split}
\label{pls_2DEG}
\end{equation}
where $\bar{\omega} = \hbar\omega/E_{F}$, $\bar{q}=q/k_{F}$, $\bar{q}_{\textsf{TF}}=q_{\textsf{TF}}/k_{F}$, and $q_{\textsf{TF}} = me^2/2\pi\hbar^2\kappa\epsilon_{0}$ is the Thomas-Fermi wavevector for the 2DEG. We note Eq. \eqref{pls_2DEG} gives valid description for $q \leq q_{c}$, the critical wave vector at which plasmon mode touches the 2DEG particle-hole continuum (PHC). The plasmon dispersion ceases to exist for $q > q_{c}$ (see Fig. 1(a)). This critical wavevector can be determined via \citep{czachor1982}: 
\begin{equation}
\begin{split}
\frac{\bar{q}^{2}_{c}}{\bar{q}_{\textsf{TF}}}+\frac{\bar{q}^{3}_{c}}{2\bar{q}^{2}_{\textsf{TF}}} = 1,
\end{split}
\label{pls_2DEG_qc}
\end{equation}  
where $\bar{q}_{c} = q_{c}/k_{F}$.\\

When a cutoff is introduced, two changes are evident (see Fig. 1(b)): (1) The PHC is reconfigured, the maximum energy for single-particle transitions is now limited by the bandwidth. (2) In addition to \textit{low-q} plasmon modes which resembles the plasmons in 2DEG, another plasmon branch appears at \textit{large-$q$}, with energies close to $E_{c}$. With decreasing the bandwidth the two branches merge, see Fig. 1(c).\\
         
These trends can be followed via a semi-analytical model, valid for clean sample ($\eta \rightarrow 0$) at zero temperature. The model is build upon the following argument: in Eq. \eqref{pol_lindhard}, for $q\leq k_{c}-k_{F}$, $E_{{\bf{k}}+{\bf{q}}} = \hbar^2|{\bf{k}}+{\bf{q}}|^{2}/2m$ and the polarizability is identical to that of a 2DEG. For $q\geq k_{c}+k_{F}$, we have $E_{{\bf{k}}+{\bf{q}}} = E_{c}$. For $ k_{c}-k_{F} \leq q \leq k_{c}$, $E_{{\bf{k}}+{\bf{q}}} = E_{c}$ only if $k_{c}-q \leq k \leq k_{F}$ and $0 \leq \theta \leq \theta_{0}$, where $\theta$ is the angle between ${\bf{k}}$ and ${\bf{q}}$, and:
\begin{equation}
\theta_{0} = \cos^{-1} \left(\frac{k^{2}_{c}-k^{2}-q^{2}}{2kq}\right),
\label{theta0}
\end{equation}
is the angle at which $|{\bf{k}}+{\bf{q}}| = k_{c}$. Likewise, for $ k_{c} \leq q \leq k_{c}+k_{F}$, $E_{{\bf{k}}+{\bf{q}}} = \hbar^2|{\bf{k}}+{\bf{q}}|^{2}/2m$, only if $q - k_{c} \leq k \leq k_{F}$ and $\theta_{0} \leq \theta \leq \pi$. Accordingly, the closed form relations for the imaginary part of the polarizability, within different phase space defined in Fig. 2(a) can be obtained as:\\  
\begin{subequations}
\begin{equation}
\begin{split}
\operatorname{Im}\,P(\bar{q}\,,\bar{\omega}) = \frac{m}{2\pi\hbar^2} \frac{1}{\bar{q}^2} \bigg( & \sqrt{4\bar{q}^2-(\bar{\omega}-\bar{q}^2)^2} \\
& ~~ - \sqrt{4\bar{q}^2-(\bar{\omega}+\bar{q}^2)^2} \bigg)
\label{imag_pol_analy_q1}
\end{split}
\end{equation}
in region i:  $0 \leq \bar{\omega} \leq \min\left\lbrace-(\bar{q}-1)^2+1,\bar{E}_{c}-1\right\rbrace$,
\begin{equation}
\operatorname{Im}\,P(\bar{q}\,,\bar{\omega}) = \frac{m}{2\pi\hbar^2} \frac{1}{\bar{q}^2}\sqrt{4\bar{q}^2-(\bar{\omega}-\bar{q}^2)^2}
\label{imag_pol_analy_q2}
\end{equation}
in region ii : $\max\left\lbrace-(\bar{q}-1)^2+1,(\bar{q}-1)^2-1\right\rbrace \leq \bar{\omega} \leq \min\left\lbrace(\bar{q}+1)^2-1,\bar{E}_{c}-1\right\rbrace$,
\begin{equation}
\begin{split}
\operatorname{Im}\,P & (q\,,\omega) = \frac{m}{2\pi\hbar^2} \frac{1}{\bar{q}^2} \bigg(\sqrt{4(\bar{E}_{c}-\bar{\omega})\bar{q}^2-(\bar{\omega}-\bar{q}^2)^2} \\
& - \sqrt{4\bar{q}^2-(\bar{\omega}+\bar{q}^2)^2}\bigg) + \frac{m}{\pi\hbar^2} \cos^{-1}\left( \frac{\bar{\omega}-\bar{q}^2}{2\bar{q}\sqrt{\bar{E}_{c}-\bar{\omega}}}\right)
\end{split}
\label{imag_pol_analy_q3}
\end{equation}
in region iii:  $\bar{E}_{c}-1 \leq \bar{\omega} \leq -(\bar{q}-1)^2+1$ (note that this region appears only if $\bar{k}_{c} \leq \sqrt{2}$),
\begin{equation}
\begin{split}
\operatorname{Im}\,P(\bar{q}\,,\bar{\omega}) = \frac{m}{2\pi\hbar^2} \frac{1}{\bar{q}^2} & \sqrt{4(\bar{E}_{c}-\bar{\omega})\bar{q}^2-(\bar{\omega}-\bar{q}^2)^2}\\
& +\frac{m}{\pi\hbar^2} \cos^{-1}\left( \frac{\bar{\omega}-\bar{q}^2}{2\bar{q}\sqrt{\bar{E}_{c}-\bar{\omega}}}\right)
\label{imag_pol_analy_q4}
\end{split}
\end{equation}
in region iv: $\max \left\lbrace -(\bar{q}-1)^2+1,\bar{E}_{c}-1 \right\rbrace\leq \bar{\omega} \leq \bar{E}_{c}-(\bar{q}-\bar{k}_{c})^{2}$, and
\begin{equation}
\operatorname{Im}\,P(\bar{q}\,,\bar{\omega}) = \frac{m}{\hbar^2}
\label{imag_pol_analy_q5}
\end{equation}
in region v: $\bar{q} \geq \bar{k}_{c}$ and $\max\left\lbrace\bar{E}_{c}-(\bar{q}-\bar{k}_{c})^{2},\bar{E}_{c}-1\right\rbrace \leq \bar{\omega} \leq \bar{E}_{c}$. We then use Kramers-Kronig transformation to obtain the real part of the polarizability. In Fig. 2(a), the boundaries of the PHC are illustrated. The plasmon dispersion obtained from this semi-analytic method compares well with the numeric result.
\label{imag_pol_analy}
\end{subequations}\\

We note in the long wavelength limit, $q\rightarrow 0$ and for a clean sample, the polarizability for FBW is identical to that of 2DEG. In fact, Eq. \eqref{pls_2DEG} also gives the plasmon modes in FBW-2DEG for $\bar{q} \leq \min\{\bar{q}_{c}\,,\,\bar{k}_{c}-1\}$ (see Fig. 2(a)). The bandwidth at which the two plasmon branches merge can be followed via: $ \bar{k}_{c}\sim \bar{q}_{c}+1$. The latter is the condition for \textit{low-q} plasmon branch to jump over PHC at $\bar{q} = \bar{k}_{c}-1$. We note that the $E_{c}$ required for merging increases with the Fermi level and decreases with $\kappa$.\\ 

\begin{figure}
\centering
\includegraphics[width=\linewidth]{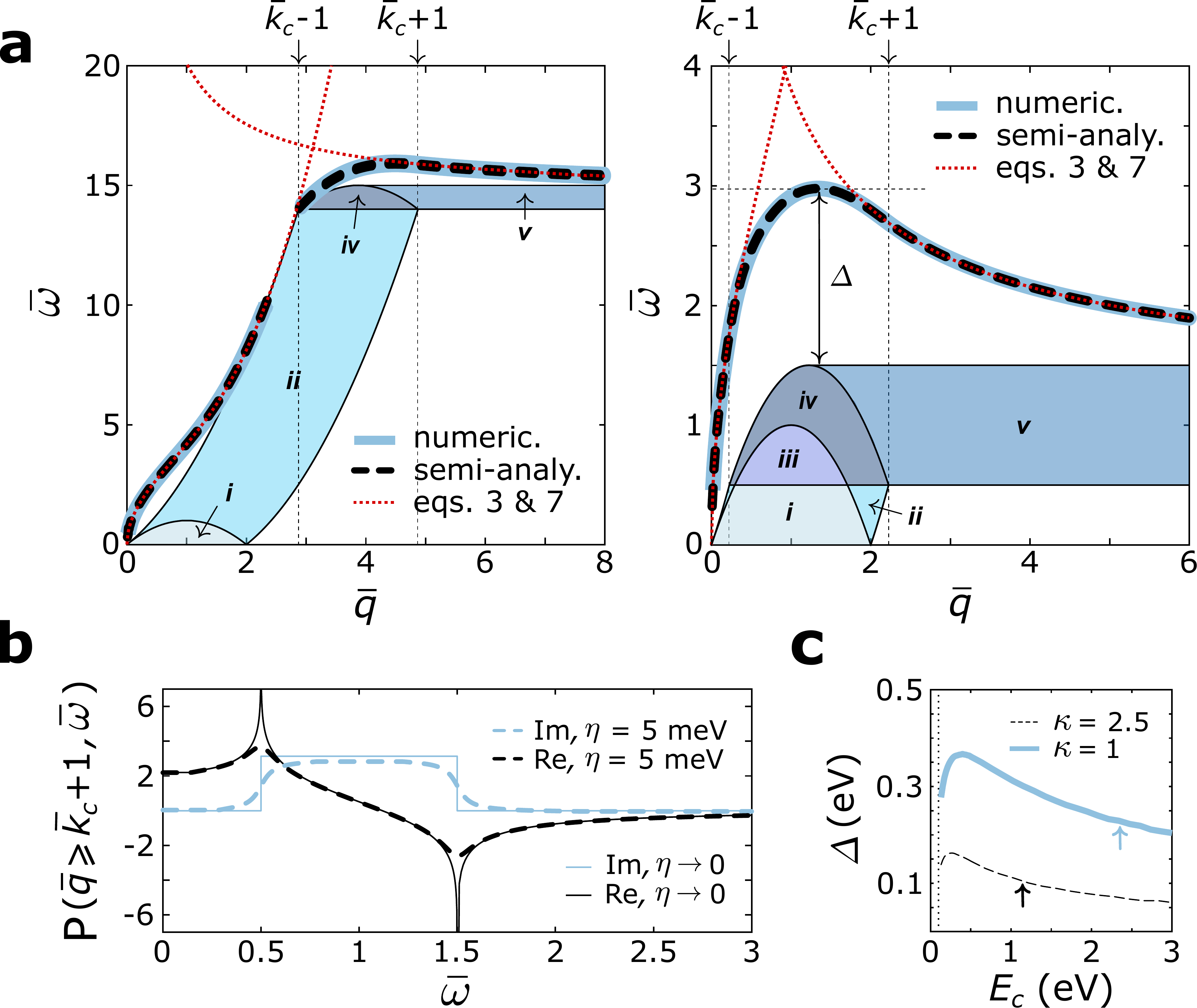}
\caption{(a) The plasmon dispersion obtained numerically and via semi-analytic model. The different regions characterizing the $\operatorname{Im} P \neq 0$ are also displayed. The left panel: $\bar{E}_{c} = 15$. The right panel: $\bar{E}_{c} = 1.5$. (b) The real and imaginary parts of the polarizability (normalized to $m/\pi\hbar^2$) for zero and 5\,meV broadening at $\bar{q} \geq \bar{k}_{c}+1$ and $\bar{E}_{c} = 1.5$. (c) $\Delta$ vs bandwidth. The dotted line is the Fermi energy. The arrows denote the bandwidth at which merging occurs.}
\label{fig2}
\end{figure}

For $\bar{q} \geq \bar{k}_{c}+1$, a closed form expression can be obtained for the plasmon dispersion in FBW-2DEG (see Fig. 2(a)):
\begin{equation}
\begin{split}
\bar{\omega}^2 = \frac{2\bar{E}_{c} - 1}{\exp\left(\bar{q}/\bar{q}_{\textsf{TF}}\right) - 1} + \bar{E}^{2}_{c}
\end{split}
\label{large_q_pls}
\end{equation}
Interestingly, although the plasmon energy is asymptotically approaching $\bar{E}_{c}$ for $\bar{q}\gg~\bar{k}_{c}+1$, at zero temperature and $\eta \rightarrow 0$, it does not enter the PHC. This can be tracked by inspecting the real part of the polarizability for $\bar{q}\geq \bar{k}_{c}+1$ which reads as:
\begin{equation}
\begin{split}
\operatorname{Re} P (\bar{q},\bar{\omega}) = \frac{m}{\pi\hbar^2}\log\left(\frac{\bar{E}^{2}_{c}-\bar{\omega}^{2}}{\left(\bar{E}_{c}-1\right)^{2}-\bar{\omega}^{2}} \right)
\end{split}
\label{large_q_pol_real}
\end{equation}
It is clear that the latter becomes singular at $\bar{\omega} = \bar{E}_{c}$, which guarantees a zero for dielectric function irrespective of the $\bar{q}$ magnitude. This, however, can be relaxed for a finite broadening, as illustrated in Fig. 2(b). For a 2D system sandwiched between two dielectric mediums, the plasmon momentum is inversely proportional to field decay length. This suggests unprecedented field confinement in FBW-2DEG.\\    

Moreover, for $q \gtrsim q_{\textsf{TF}} \log\left(E_{F}/\Delta E\right)$, the plasmon energy varies in the range $~\Delta E$ above $E_{c}$, which allows for plasmon modes with almost constant dispersion to emerge. The onset $q$ for this quasi-flat dispersion can also be tuned with the dielectric choice or Fermi energy. This allows for tunable near-field exponential amplification in a setup which has two FBW-2DEGs separated by a dielectric spacer \citep{stauber2012, stauber2016}.\\  

We next focus on the plasmon mode with maximum energy, where we introduce $\Delta$ denoting its energy relative to the upper boundary of the PHC, see Fig. 2(a). The importance of this mode is two-fold: (1) Low-loss plasmon. This mode is arguably protected against Landau damping mediated by elastic scattering, since pure momentum transfer will not be adequate to bring it into the PHC. Moreover, given $\Delta \gg E_{c}$, damping pathways due to optical phonon inelastic scattering are also quenched. We note the idea to achieve lossless plasmons in narrow band electronic systems has been explored in the context of bulk metals via computing the local dielectric function \citep{gjerding2017, khurgin2010}. The local description, by definition assumes identical loss for all momenta, and gives an incomplete view of the PHC boundaries. It thus cannot be used to predict the mode robustness against Landau-damping, an intrinsic major damping channel in 2D metals \citep{hugen2013}. In Fig. 2(c), we show in FBW-2DEG, dielectric engineering is a plausible route to protect maximum energy plasmons against both elastic and inelastic scattering-mediated Landau damping. Also note that $\Delta$ vs bandwidth has a maximum which occurs at $E_{c}$ well below the bandwidth required for merging and larger than Fermi energy. (2) Negative group velocity. We note group velocity changes sign at the momentum pertaining to maximum energy plasmon. This mode occurs at $q\sim k_{c}$ which implies the onset for negative dispersion can be tuned via controlling the bandwidth. The tunablity of the group velocity sign allows for interesting phenomena such as all-angle negative refraction, normal Doppler frequency shift, or tunable directional plasmon excitation \citep{agranovich2006, shin2006, lin2019, jiang2018}.\\

\begin{figure*}
\centering
\includegraphics[width=0.85\linewidth]{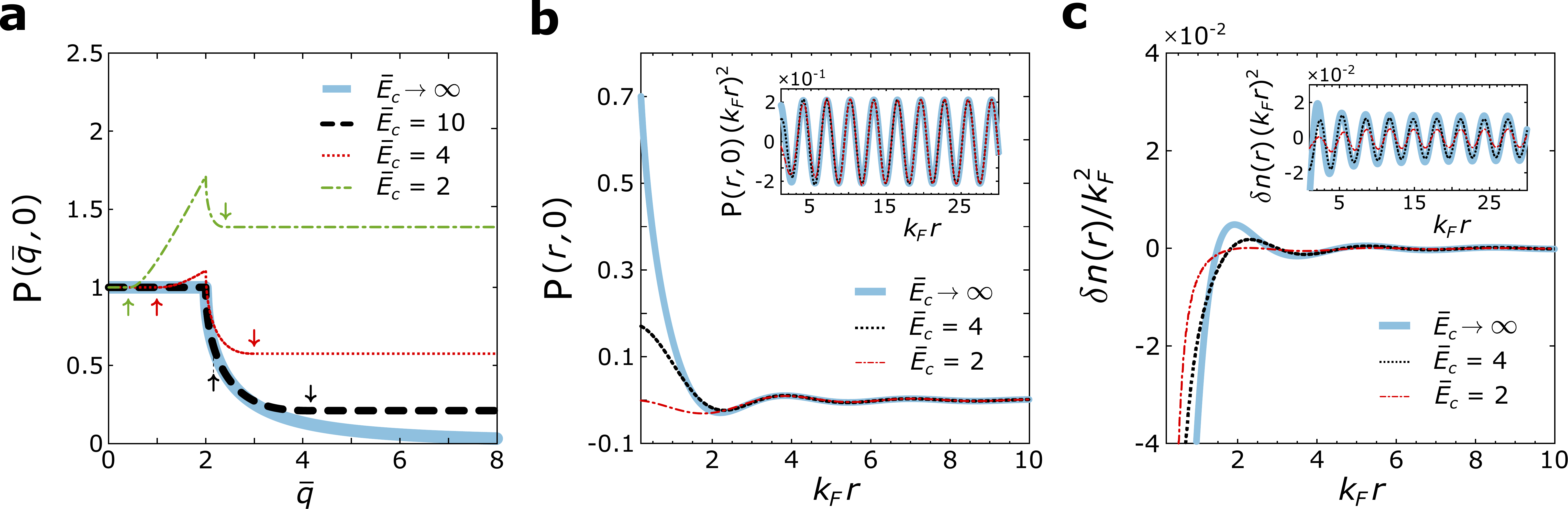}
\caption{(a) The static polarizability (normalized to $m/\pi\hbar^2$) vs $\bar{q}$. The upward (downward) arrows denote $\bar{q} = \bar{k}_c-1$ ($\bar{q} = \bar{k}_c+1$). (b) The spatial Fourier transform of the static polarizability (normalized to $m\,k^{2}_{F}/\pi\hbar^2$) as a function of $k_{F}r$. The inset shows the $r^{-2}$ decay at large distances. (c) same as (b) for the induced charge density $\delta n(r)$ (in units of $k^{2}_{F} Ze$).}
\label{fig3}
\end{figure*}

\noindent \textit{Static Screening \& Friedel Oscillations} -- The static limit of the polarizability, $\omega = 0$ with arbitrary $q$ is relevant for the screening of charged and magnetic impurities. In Fig. 3(a), we cover this limit. For a 2DEG with $E_{c} \rightarrow \infty$, the static polarizability is given by \citep{stern1967}:     
\begin{equation}
\begin{split}
P(\bar{q},0) = \frac{m}{\pi\hbar^2} \left( 1-\theta(\bar{q}-2)\sqrt{1 - \frac{4}{\bar{q}^2}}\,\right)
\end{split}
\label{static_pol_2DEG}
\end{equation}
This is also valid for FBW-2DEG given that $\bar{q} \leq \bar{k}_{c}-1$. For 2DEG at short-wavelength limit the static polarizability falls off rapidly (with $1/\bar{q}^{2}$). For FBW-2DEG, however, given that $\bar{q} \geq \bar{k}_{c}+1$ the static polarizability is given by: 
\begin{equation}
\begin{split}
P(\bar{q},0) = \frac{2m}{\pi\hbar^2} \log\left(\frac{\bar{E}_{c}}{\bar{E}_{c}-1}\right),
\end{split}
\label{static_pol_FBW_largeq}
\end{equation}
i.e. a constant value which depends on the bandwidth and Fermi energy. Moreover, we note that similar to the 2DEG, the polarizability for FBW also has a kink at $\bar{q} = 2$, although its magnitude at $\bar{q} = 2$ increases from that of 2DEG given $\bar{k}_{c} \leqslant 3$. It is worthy to mention the increase beyond 2DEG density of states in the static polarizability for $\bar{k}_{c}-1 \leq \bar{q} \leq 2$ can be attributed to smaller energy denominator in Eq. \eqref{pol_lindhard} at $\omega = 0$, which is now limited by the bandwidth. We note in passing that the polarizability kink also implies an observable Kohn anomaly in FBW-2DEG vibration spectrum \citep{kohn1959}.\\  

We next discuss the Fourier transform of the static polarizability, given by:
\begin{equation}
\begin{split}
P({\bf{r}},0) = \frac{1}{4\pi^2} \int d^{2}q\,P(q,0)e^{i{\bf{q}}.{\bf{r}}},
\end{split}
\label{Pr}
\end{equation}
which determines the induced spin density at position $\bf{r}$ due a magnetic impurity located at the origin \citep{fischer1975}. It also gives the Ruderman-Kittel-Kasuya-Yosida (RKKY) interaction energy between two magnetic impurities, where each impurity interacts with the density induced by the other \citep{beal1987}. From Fig. 3(b), the finite bandwidth does not affect the Friedel oscillations which appear in the induced spin density (neither the period nor the amplitude) at large distances ($k_{F}r\gg 1$). However, close to magnetic impurity the magnitude of the induced spin density is strongly decreased.\\

We consider next an external charge $n_{\textsf{ext}}(r) = Ze\,\delta(r)$ screened by FBW-2DEG, which results in an induced charge density $\delta n(r)$ \citep{giuliani2005}:
\begin{equation}
\begin{split}
\delta n(r) = \frac{Ze}{4\pi^2} \int d^{2}q\left(\frac{1}{\epsilon(q,0)}-1\right)e^{i{\bf{q}}.{\bf{r}}},
\end{split}
\label{dnr}
\end{equation}
According to Fig. 3(c), the induced charge density for FBW follows the same form as the 2DEG, with an oscillatory behavior (with wavelength $ \sim \pi/k_{F}$), which oscillates around zero and decays with $1/r^{2}$ at large distances, as verified in the inset of Fig. 3(c). The oscillation amplitude, however, strongly decreases for narrower bandwidth.\\

\begin{figure*}
\centering
\includegraphics[width=0.75\linewidth]{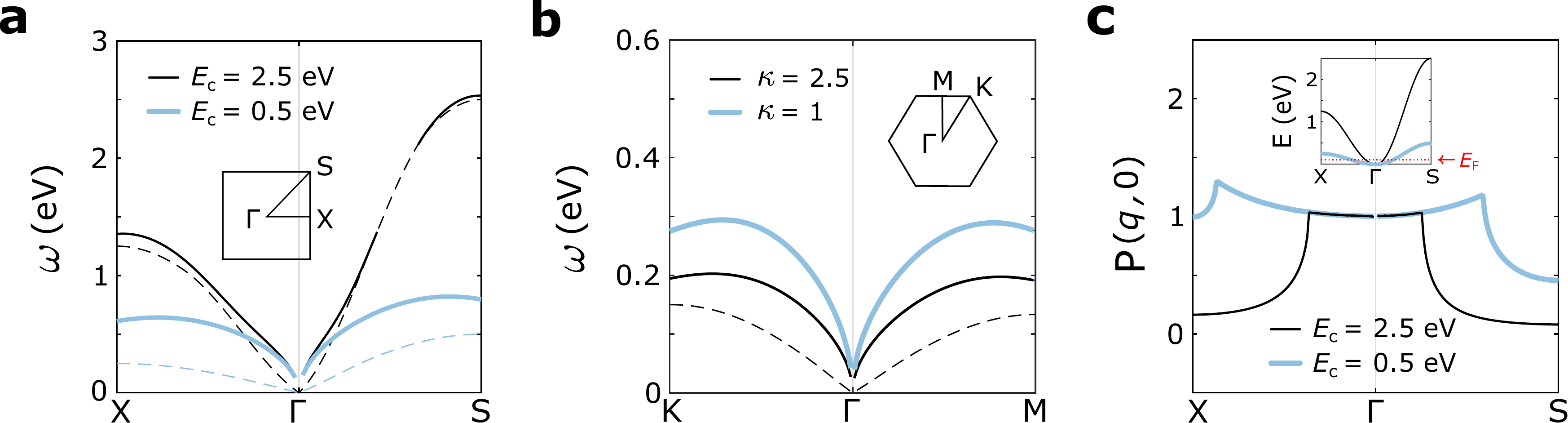}
\caption{(a) The plasmon dispersion for the square lattice with $E_{c} = 2.5$\,eV and 0.5\,eV. (b) The plasmon dispersion for hexagonal lattice with $E_{c} = 0.15$\,eV and $\kappa = 2.5$ and 1. The dashed lines in (a) and (b) denote the upper boundary of the PHC. (c) The static polarizability (normalized to density of states at $E_{F}$) of the square lattice for $E_{c} = 2.5$\,eV and 0.5\,eV, with the electron energy dispersions depicted in the inset. In all panels, $E_{F} = 0.1$\,eV and $a = 4$\,\AA.}
\label{fig4}
\end{figure*}

Before closing, we note that the presence of the cusp at $k_{c}$ or the extended flat band at large momenta would not change the overall qualitative picture, and the main attributes predicted by the toy model remain valid even with these relaxed. This can be confirmed by computing the polarizability for a simple hexagonal or square lattice with one orbital per basis which gives a single electronic band in tight binding approximation. Accordingly, for the square lattice we have: $E_{\bf{k}} = -2 \gamma \left[ \cos(k_{x}a) + \cos(k_{y}a) \right]$, and for the hexagonal lattice: $E_{\bf{k}} = -2 \gamma \left[\cos(k_{x}a) + 2\cos(k_{x}a/2)\cos(k_{y}a\sqrt{3}/2) \right]$, where $a$ and $\gamma$ denote the lattice constant and the nearest neighbor hopping amplitude, respectively. To give a fixed bandwidth $E_{c}$, we set $\gamma$ to $E_{c}/8$ ($E_{c}/9$) for the square (hexagonal) lattice. The panels in Fig. 4, clearly show that the salient features predicted by our simple model can be reproduced with these tight-binding models. This includes the plasmonic branches at large $E_{c}$, their merging for narrower bandwidths, the negative group velocity appears close to zone edge, the increase in $\Delta$ with $\kappa$, the kink in the static polarizability and its evolution with the bandwidth.\\ 


\noindent \textit{Conclusion} -- In summary, we compute the dielectric function of a 2DEG with finite bandwidth. The dispersing plasmons include modes which can arguably be immune to Landau damping mediated by elastic and inelastic scattering. Moreover, the dispersion has a quasi-flat tail which can be extended to large momenta. In the static limit, the FBW polarizability differs from the 2DEG, especially in the large momenta limit, where it saturates to a constant value as opposed to 2DEG where it falls rapidly to zero. Moreover, the static Friedel charge and spin density oscillations due to a charged or magnetic impurity embedded in FBW-2DEG are recorded. The non-analyticity at $q = 2k_{F}$ leads to Friedel oscillations with period $\sim \pi/k_{F}$ decaying as $r^{-2}$ at large distances. The oscillation amplitudes in induced charge increases with the bandwidth, while the the amplitude remain intact in the induced spin density.\\  
  
\noindent \textit{Note added} - Recently, we became aware of Ref. \citep{lewandowski2019} in which the authors discuss low-damped plasmonic modes in narrow band electronic systems with the focus on twisted bilayer graphene.\\

\noindent \textit{Acknowledgments} -- This work has been supported by Spain's MINECO under Grant No. FIS2017-82260-P as well as by the CSIC Research Platform on Quantum Technologies PTI-001. K.K. and T.L. acknowledge support by the National Science Foundation NSF/EFRI grant ($\#$EFRI-1741660). 


\end{document}